\documentstyle[aps,twocolumn,epsf]{revtex}
\newcommand{\be}{\begin{equation}}
\newcommand{\ee}{\end{equation}}
\newcommand{\bea}{\begin{eqnarray}}
\newcommand{\eea}{\end{eqnarray}}

\newlength{\charwidth}
\def\medhat#1{\settowidth{\charwidth}{$#1\,$}{\makebox[\charwidth]{$\,
 {\widehat{\makebox[2mm]{$#1\,$}}}$}}\vphantom{#1}}
\def\vhight#1{\vphantom{\left(\begin{picture}(0,#1)\end{picture}\right)}
}
\def\DSa{\begin{picture}(0,0)\thicklines
\put(0,0){\oval(1.5,1)}
\put(0,0){\makebox(0,0){$-\ii\Sa$}}\end{picture}}
\def\GlnG0Sa{
\begin{picture}(4.3,1.5)
\put(0.5,.1){
\put(.5,0){\DSa}\put(3,0){\DSa}\put(1.75,1){\DSa}
\put(1.75,-1){\makebox(0,0){\dots\dots}}
\put(1,.5){\oval(1,1)[lt]}\put(2.5,.5){\oval(1,1)[rt]}
\put(1,-.5){\oval(1,1)[lb]}\put(2.5,-.5){\oval(1,1)[rb]}}
\end{picture}}
\def\GGaSa{
\begin{picture}(2.5,2.)
\thicklines\put(0.5,.1){
\put(.5,0){\DSa}\put(2,-.5){\line(0,1){1}}
\put(1.,1){\line(1,0){0.5}}\put(1.,-1){\line(1,0){0.5}}
\put(1,.5){\oval(1,1)[lt]}\put(1.5,.5){\oval(1,1)[rt]}
\put(1,-.5){\oval(1,1)[lb]}\put(1.5,-.5){\oval(1,1)[rb]}}
\end{picture}}
\def\Dclosed#1#2{
\begin{picture}(2,.8)\put(0,.1){#2
\put(1,0){\circle{1.414}}
\put(1,-.707){\line(0,1){1.414}}\put(.293,0){\line(1,0){1.414}}
\put(0.5,-0.5){\line(0,1){1}}\put(1.5,-0.5){\line(0,1){1}}
\put(0.5,-0.5){\line(1,0){1}}\put(0.5,0.5){\line(1,0){1}}}
\put(1.5,-.8){$#1$}
\end{picture}}

\def\pls{\makebox(0,0){$+$}}

\def\ssp{\makebox(0,0)
    {\thinlines\put(-.1,0){\line(1,0){.2}}\put(0,-.1){\line(0,0){.2}}}}
\def\ssm{\makebox(0,0){\put(-.1,0){\thinlines\line(1,0){.2}}}}
\def\photon{\thinlines\multiput(0,0)(.2,0){3}{\line(1,0){0.1}}}

\def\Photon{\thicklines\multiput(0,0)(0.2,0){5}{\line(1,0){0.1}}
\put(0.4,0){\vector(1,0){.2}}}

\def\VFermion{\thicklines\put(0,0){\vector(0,1){.8}}
            \put(0,.8){\line(0,1){.5}}}
\def\AVFermion{\thicklines\put(0,0){\vector(0,-1){.8}}
            \put(0,-.8){\line(0,-1){.5}}}

\def\oneloop{
     \put(1.5,0){\thicklines\oval(2.0,1.5)}
     \put(0,0){\photon}\put(0.3,0.3){\ssp} 
     \put(2.5,0){\photon}\put(2.7,0.3){\ssm}}

\def\hatchedself{\begin{picture}(5,1)\put(0,0){\oneloop}
     \multiput(0.95,0)(.2,0){5}{\put(0,-.7){\line(1,4){.35}}
     \put(0,.7){\line(1,-4){.35} }}
     \end{picture}}
\def\oneloopvertex{
    \put(1.625,0){\thicklines\oval(2.0,1.5)}
    \put(0,0){\photon}\put(0.35,0.3){\ssp} 
    \put(0.625,0){\circle*{.25}}\put(2.625,0){\circle*{.25}}
    \put(2.75,0){\photon}\put(2.9,0.3){\ssm}}

\def\longloop{
             \put(3.125,0){\thicklines\oval(5.0,1.5)}
             \put(0,0){\photon}\put(0.35,0.3){\ssp} 
             \put(5.75,0){\photon}\put(5.9,0.3){\ssm}
             \put(0.625,0){\circle*{.25}}\put(5.625,0){\circle*{.25}}}

\def\fullbox{\makebox(0,0){\rule{1.5mm}{3mm}}}
\def\interaction{\makebox(0,0){\put(0,0){\interact}
    \put(0,.95){\ssp}\put(0,-.95){\ssm}
    \put(0,.125){\ssp}\put(0,-.125){\ssm}}}
\def\interact{\makebox(0,0){\put(0,.5){\fullbox}
    \thicklines\put(0,0){\oval(.75,.5)}
    \put(0,-.5){\fullbox}} }

\def\til2loop{\put(0,0){\oneloopvertex}\put(4.,0){\pls}
      \put(4.75,0){\oneloopvertex}\put(6.375,0){\interaction}
      \put(9,0){\pls}}
\def\classdiagram{
     \begin{picture}(12,6)
     \put(0,2.5){\til2loop}\put(2,0){\nloop}
     \put(9,0){\pls}\put(10,0){\makebox(0,0){$\dots$}}
     \put(10,2.5){\makebox(0,0){$\dots$}}  
     \end{picture}}

\def\nloop{\begin{picture}(6.5,.5) \put(0,0){\longloop}
       \multiput(1.625,0)(1,0){2}{\interaction}\put(3.625,0)
       {\makebox(0,0){$\dots$}}\put(4.625,0){\interaction}\end{picture}}

\def\contourxy{\unitlength=0.445cm
\begin{picture}(17,1)\thicklines
\contour
\put(14.5,-.5){\makebox(0,0){$\infty$}}
\put(11,-.5){\makebox(0,0){$t_x^+$}}
\put(8,1.8){\makebox(0,0){$t_y^-$}}
\put(11,0){\circle*{.2}}
\put(8,1){\circle*{.2}}
\end{picture}}

\def\contour{\thicklines
\put(1,-.5){\makebox(0,0){$t_{0}$}}
\put(16.5,.5){\makebox(0,0){$t$}}
\put(0,.5){\vector(1,0){16}}
\put(1,1.){\line(1,0){13}}\put(14,.5){\oval(1,1)[br]}
\put(14,0){\vector(-1,0){13}}\put(14,.5){\oval(1,1)[tr]}}

\newcommand{\di}{{\rm d}}
\newcommand{\ii }{{\rm i}}

\def\scr#1{\mbox{\scriptsize #1}}\def\Do{{\cal D}}\def\vu{v} 
\def\Lg{{\cal L}}
\def\Lgh{\makebox[3.5mm]{${\widehat{\makebox[2mm]{$\Lg$}}}$}\vphantom{L}}
\def\Lint{\Lgh^{\mbox{\scriptsize int}}}
\def\dpi#1{\frac{\di^4 #1}{(2\pi)^4}}                
\def\Pbr#1{\left\{#1\right\}}                    
\def\Gr{G}\def\Ga{G}\def\Se{\Sigma}\def\Sa{\Sigma}
\def\A{A}
\def\Gm{\Gamma}
\def\F{F}                             
\def\Ft{\widetilde{F}}                 
\def\Fd{F}                             
\def\Fdt{\widetilde{F}}                
\def\fd{f}

\def\Get{\Gamma_{\scr{out}}}   
\def\Gbt{\Get}                                   
\def\Ge{\Gamma_{\scr{in}}}     
\def\Gb{\Ge}              

\def\Ld{\Gamma_{\scr{out}}}
\def\Ldt{\Gamma_{\scr{in}}}
\def\Re{\mbox{Re}\;}\def\Im{\mbox{Im}\;}
\def\loopa{
\unitlength 1.00mm
\thicklines
\begin{picture}(10.00,8.93)
\bezier{28}(2.00,5.00)(2.13,8.33)(6.00,8.93)
\bezier{28}(2.00,4.87)(2.13,1.53)(6.00,0.93)
\bezier{28}(10.00,5.00)(9.87,8.33)(6.00,8.93)
\bezier{28}(10.00,4.87)(9.87,1.53)(6.00,0.93)
\bezier{28}(2.00,-3.00)(2.13,0.33)(6.00,0.93)
\bezier{28}(2.00,-3.13)(2.13,-6.47)(6.00,-7.07)
\bezier{28}(10.00,-3.00)(9.87,0.33)(6.00,0.93)
\bezier{28}(10.00,-3.13)(9.87,-6.47)(6.00,-7.07)
\put(6.00,1.00){\circle*{1.00}}
\put(6.40,8.80){\vector(1,0){0.2}}
\put(5.53,8.73){\line(1,0){0.87}}
\put(5.53,-6.93){\vector(-1,0){0.2}}
\put(6.47,-7.00){\line(-1,0){0.93}}
\end{picture}}
\def\loopb#1{
\unitlength 1.00mm
\thicklines
\begin{picture}(17.50,13.00)
\put(2.00,1.00){\circle*{1.00}}
\put(17.00,1.00){\circle*{1.00}}
\bezier{68}(2.00,1.00)(9.50,5.00)(17.00,1.00)
\bezier{68}(2.00,1.00)(9.50,-3.00)(17.00,1.00)
\bezier{112}(2.00,1.00)(9.50,13.00)(17.00,1.00)
\bezier{112}(2.00,1.00)(9.50,-11.00)(17.00,1.00)
\put(9.93,-1.07){\vector(1,0){0.2}}
\put(9.00,-1.07){\line(1,0){0.93}}
\put(9.93,2.93){\vector(1,0){0.2}}
\put(8.93,2.93){\line(1,0){1.00}}
\put(9.13,6.93){\vector(-1,0){0.2}}
\put(10.07,6.93){\line(-1,0){0.93}}
\put(9.20,-5.00){\vector(-1,0){0.2}}
\put(10.07,-5.07){\line(-1,0){0.87}}
\if#1c \multiput(9.5,-8.)(0,6.5){3}{\thinlines\line(0,1){5}}\fi
\end{picture}}
\def\loopc#1{ 
\unitlength 1.00mm
\thicklines
\begin{picture}(18.54,9.47)
\put(2.00,-5.00){\circle*{0.93}}
\put(18.00,-5.00){\circle*{1.07}}
\put(10.00,9.00){\circle*{0.94}}
\bezier{72}(2.00,-5.00)(10.00,-1.00)(18.00,-5.00)
\bezier{72}(2.00,-5.00)(10.00,-9.00)(18.00,-5.00)
\bezier{80}(2.00,-5.00)(2.00,6.00)(10.00,9.00)
\bezier{72}(2.00,-5.00)(9.07,-0.80)(10.00,9.00)
\bezier{72}(18.00,-5.00)(11.07,-0.93)(10.00,9.00)
\bezier{76}(18.00,-5.00)(18.13,5.07)(10.00,9.00)
\put(9.93,-7.00){\vector(-1,0){0.2}}
\put(11.00,-7.00){\line(-1,0){1.07}}
\put(3.80,3.27){\vector(1,2){0.2}}
\multiput(3.40,2.47)(0.10,0.20){4}{\line(0,1){0.20}}
\put(16.73,2.33){\vector(2,-3){0.2}}
\multiput(16.20,3.13)(0.11,-0.16){5}{\line(0,-1){0.16}}
\put(10.07,-3.00){\vector(1,0){0.2}}
\put(9.07,-3.00){\line(1,0){1.00}}
\put(12.53,0.60){\vector(-2,3){0.2}}
\multiput(13.00,-0.20)(-0.12,0.20){4}{\line(0,1){0.20}}
\put(7.60,0.67){\vector(-2,-3){0.2}}
\multiput(8.00,1.33)(-0.10,-0.17){4}{\line(0,-1){0.17}}
\if#1c \multiput(8.5,-9.5)(3.5,5.25){3}{\thinlines\line(2,3){2.5}}\fi
\end{picture}}
\def\wa#1{
\unitlength 0.7mm
\thicklines
\begin{picture}(7,5)\put(2,0){
\put(2.00,1.00){\circle*{1.00}}
\put(2.00,6.00){\makebox(0,0){$ #1 $}}
\bezier{16}(0.54,-0.41)(3.54,2.55)(3.54,2.55)
\bezier{16}(0.54,2.55)(3.50,-0.41)(3.50,-0.41)
\put(1.21,0.22){\vector(1,1){0.2}}
\multiput(0.54,-0.45)(0.11,0.11){6}{\line(0,1){0.11}}
\put(4.61,3.51){\vector(1,1){0.8}}
\multiput(2.32,1.22)(0.11,0.11){20}{\line(0,1){0.11}}
\put(1.25,1.76){\vector(3,-4){0.2}}
\multiput(0.58,2.55)(0.11,-0.13){6}{\line(0,-1){0.13}}
\put(4.54,-1.37){\vector(1,-1){0.8}}
\multiput(2.25,0.76)(0.13,-0.12){18}{\line(1,0){0.13}}
}
\end{picture}}
\def\wb#1{
\unitlength 0.7mm
\thicklines
\begin{picture}(17.51,9.01)\put(4,0){
\put(9.51,-6.49){\circle*{1.00}}
\put(9.51,8.51){\circle*{1.00}}
\put(9.51,-9.49){\makebox(0,0){$ #1 $}}
\put(9.51,11.51){\makebox(0,0){$ #1 $}}
\bezier{68}(9.51,-6.49)(5.51,1.01)(9.51,8.51)
\bezier{68}(9.51,-6.49)(13.51,1.01)(9.51,8.51)
\put(11.58,1.44){\vector(0,1){0.2}}
\put(11.58,0.51){\line(0,1){0.93}}
\put(7.52,0.58){\vector(0,-1){0.2}}
\put(7.52,1.51){\line(0,-1){0.93}}
\bezier{20}(7.00,8.55)(12.04,8.55)(12.04,8.55)
\bezier{20}(7.00,-6.49)(12.04,-6.53)(12.04,-6.53)
\put(8.00,8.51){\vector(1,0){0.2}}
\put(7.00,8.51){\line(1,0){1.00}}
\put(13.44,8.55){\vector(1,0){0.2}}
\put(12.44,8.55){\line(1,0){1.00}}
\put(8.00,-6.58){\vector(1,0){0.2}}
\put(7.00,-6.58){\line(1,0){1.00}}
\put(13.44,-6.58){\vector(1,0){0.2}}
\put(12.44,-6.58){\line(1,0){1.00}}
}
\end{picture}
}
\def\pia{
\unitlength 1.00mm
\thicklines
\begin{picture}(10.00,8.93)
\bezier{28}(2.00,5.00)(2.13,8.33)(6.00,8.93)
\bezier{28}(2.00,4.87)(2.13,1.53)(6.00,0.93)
\bezier{28}(10.00,5.00)(9.87,8.33)(6.00,8.93)
\bezier{28}(10.00,4.87)(9.87,1.53)(6.00,0.93)
\put(6.00,1.00){\circle*{1.00}}
\put(3.93,1.00){\vector(1,0){0.2}}
\put(2.93,1.00){\line(1,0){1.00}}
\put(6.60,8.80){\vector(1,0){0.2}}
\put(5.60,8.80){\line(1,0){1.00}}
\put(9.83,1.00){\vector(1,0){0.2}}
\put(3.90,1.00){\line(1,0){5.93}}
\end{picture}}
\def\pib{
\unitlength 1.00mm
\thicklines
\begin{picture}(19.78,13.00)
\put(2.00,1.00){\circle*{1.00}}
\put(17.00,1.00){\circle*{1.00}}
\bezier{112}(2.00,1.00)(9.50,13.00)(17.00,1.00)
\put(9.47,1.00){\vector(1,0){0.2}}
\bezier{28}(2.00,1.00)(9.47,1.00)(9.47,1.00)
\put(9.47,6.93){\vector(-1,0){0.2}}
\put(10.07,6.93){\line(-1,0){0.60}}
\bezier{76}(2.00,1.00)(9.60,6.47)(17.00,1.00)
\put(10.00,3.60){\vector(1,0){0.2}}
\put(9.00,3.60){\line(1,0){1.00}}
\bezier{8}(-0.07,1.00)(1.93,1.00)(1.93,1.00)
\put(0.60,1.07){\vector(1,0){0.2}}
\put(-0.00,1.07){\line(1,0){0.60}}
\put(19.78,1.00){\vector(1,0){0.2}}
\put(9.44,1.00){\line(1,0){10.33}}
\end{picture}}
\def\pic{
\unitlength 1.00mm
\thicklines
\begin{picture}(21.78,15.60)
\put(2.07,1.13){\circle*{0.93}}
\put(18.07,1.13){\circle*{1.07}}
\put(10.07,15.13){\circle*{0.94}}
\bezier{80}(2.07,1.13)(2.07,12.13)(10.07,15.13)
\bezier{72}(2.07,1.13)(9.14,5.33)(10.07,15.13)
\bezier{72}(18.07,1.13)(11.14,5.20)(10.07,15.13)
\bezier{76}(18.07,1.13)(18.20,11.20)(10.07,15.13)
\put(3.87,9.40){\vector(1,2){0.2}}
\multiput(3.47,8.60)(0.10,0.20){4}{\line(0,1){0.20}}
\put(16.80,8.46){\vector(2,-3){0.2}}
\multiput(16.27,9.26)(0.11,-0.16){5}{\line(0,-1){0.16}}
\put(12.60,6.73){\vector(-2,3){0.2}}
\multiput(13.07,5.93)(-0.12,0.20){4}{\line(0,1){0.20}}
\put(7.67,6.80){\vector(-2,-3){0.2}}
\multiput(8.07,7.46)(-0.10,-0.17){4}{\line(0,-1){0.17}}
\put(10.07,1.13){\vector(1,0){0.2}}
\bezier{32}(2.07,1.13)(10.07,1.13)(10.07,1.13)
\bezier{8}(-0.00,1.00)(1.00,1.00)(2.00,1.00)
\put(0.73,1.00){\vector(1,0){0.2}}
\put(-0.07,1.00){\line(1,0){0.80}}
\put(21.78,1.11){\vector(1,0){0.2}}
\put(10.22,1.11){\line(1,0){11.56}}
\end{picture}}
\begin{document}
\title{
Towards a Quantum Transport Description of Particles\\ 
with finite Mass Width$^1$}

\author{J\"orn Knoll, GSI, Darmstadt}
\address{J.Knoll@gsi.de;
    http://theory.gsi.de} 
\author{Yu. B. Ivanov, Kurchatov Institute, Moscow}
\author{D. N. Voskresensky, Moscow Institute for Physics and Engineering}
\date{Sept. 15, 1998}
\maketitle
\begin{abstract}
  The effects of the propagation of particles which have a finite life time
  and an according width in their mass spectrum are discussed in the context
  of transport descriptions. In the first part the coupling of soft photon
  modes to a source of charged particles is studied in a classical model which
  can be solved completely in analytical terms. The solution corresponds to a
  re-summation of certain field theory diagrams. The second part addresses the
  derivation of transport equations which also account for the damping width
  of the particles. The $\Phi$-derivable method of Baym is used to derive a
  self-consistent and conserving scheme. For this scheme a conserved
  energy-momentum tensor can be constructed. Furthermore, a kinetic entropy
  can be derived which besides the standard quasi-particle part also includes
  contributions from fluctuation.
\end{abstract}
\footnotetext[1]{Talk presented by J. K. on the 4$^{th}$ Thermal Field
  Theory Workshop, Regensburg, Aug. 10 - 15, 1998}
\section{Introduction}
With the aim to describe the collision of two nuclei at intermediate or even
high energies one is confronted with the fact that the dynamics has to include
particles like the delta or rho-meson resonances with life-times of less than
2 fm/c or equivalently with damping widths above 100 MeV. The collision rates
deduced from presently used transport codes are comparable in magnitude,
whereas typical mean kinetic energies as given by the temperature range
between 70 to 150 MeV depending on beam energy. Thus, the damping width of
most of the constituents in the system can no longer be treated as a
perturbation.

As a consequence the mass spectrum of the particles in the dense matter is no
longer a sharp delta function but rather acquires a width due to collisions
and decays. The corresponding quantum propagators $G$ (Green's functions) are
no longer the ones as in the standard text books for fixed mass, but rather
have to be folded over a spectral function $A(\epsilon,{\vec p})$, which takes
a Lorentz shape $A(\epsilon,{\vec p})\sim\Gamma/((\epsilon-\epsilon({\vec
  P}))^2+(\Gamma/2)^2)$ of width $\Gamma/2$ in simple approximations. One thus
comes to a picture which unifies {\em resonances} which have already a width
in vacuum due to decay modes with the ''states'' of particles in dense matter,
which obtain a width due to collisions (collisional broadening). The
theoretical concepts for a proper many body description in terms of a real
time non equilibrium field theory have already been devised by Schwinger,
Kadanoff, Baym and Keldysh\cite{SKBK} in the early sixties, extensions to
relativistic plasmas by Bezzerides and DuBois \cite{BB}. First
investigations of the quantum effects on the Boltzmann collision term were
given Danielewicz\cite{D}, the principal conceptual problems on the level of
quantum field theory were investigated by Landsmann\cite{Landsmann}, while
applications which seriously include the finite width of the particles in
transport descriptions were carried out only in recent times,
e.g.\cite{D,DB,BM,HFN,PH,QH,Weinhold,KV}.  For resonances, e.g. the delta
resonance, it was natural to consider broad mass distributions and ad hoc
recipes have been invented to include this in transport simulation models.
However, many of these recipes are not correct as they violate some basic
principle like detailed balance\cite{DB}, and the description of resonances in
dense matter has to be improved\cite {Weinhold}.

In this talk the consequences of the propagation of particles with
short life times is re-addressed and discussed. In the first part a
genuine soft mode problem is studied: the coupling of a coherent
classical field, the Maxwell field, to the stochastic Brownian motion
of a charged particle. The rate of photons due to Bremsstrahlung,
given by the classical current-current correlation function, can be
obtained in closed analytical terms and discussed as a function of the
macroscopic transport properties, the friction and diffusion
coefficient of the Brownian particle. The result corresponds to a
partial re-summation of photon self energy diagrams in the real-time
formulation of field theory. In the second part of this talk a scheme
is presented, how to come to a self-consistent, conserving and
thermodynamically consistent transport description of particles with
finite mass width within the real-time formulation of non-equilibrium
field theory.  The derivation is based on and generalizes the
$\Phi$-functional method of Baym \cite{Baym}.  The first-order
gradient approximation provides a set of coupled equations of
time-irreversible generalized kinetic equations for the slowly varying
space-time part of the phase-space distributions and retarded
equations, which provides the fast micro-scale dynamics represented by
the four-momentum part of the distributions.  Functional methods
permit to derive a conserved energy-momentum tensor which also
includes corrections arising from fluctuations besides the standard
quasi-particle terms. Memory effects \cite{CGreiner} appearing in
collision term diagrams of higher order are discussed.  The
variational properties of $\Phi$-functional permit to derive a
generalized expression for the non-equilibrium kinetic entropy flow,
which includes corrections from fluctuations and memory effects. In
special cases we demonstrate that the entropy can only increase with
time ($H$-theorem).

\section{Preliminaries}
The standard text-book transition rate in terms of Fermi's golden
rule, e.g. for the photon radiation from some initial
state $\left|i\right>$ with occupation $n_i$ to final states
$\left|f\right>$
\def\decay{\put(1,-1.1){\VFermion}\put(0.5,-.8){\makebox(0,0){$i$}}
\put(0.5,1.2){\makebox(0,0){$f$}}
\put(1,0.2){\VFermion}\put(1,0.2){\Photon}}
\def\decayA{\put(1.5,0.2){\AVFermion}\put(2,-.8){\makebox(0,0){$i$}}
\put(2,1.2){\makebox(0,0){$f$}}
\put(1.5,1.5){\AVFermion}\put(.5,.2){\Photon}}
\begin{eqnarray}\label{Wif}\unitlength5mm
W=&\sum_{if}&n_i(1-n_f)\;\unitlength5mm
\left|\begin{picture}(2.5,1.5)\decay\end{picture}\right|^{\;\mbox{2}}\;
\nonumber\\[1ex]
&&\times(1+n_\omega)\;\delta(E_i-E_f-\omega_{\vec q})
\end{eqnarray}
with occupation $n_{\omega}$ for the photon, is limited to the concept of
asymptotic states. It is therefore inappropriate for problems which deal with
particles of finite life time. One rather has to go to the ``closed'' diagram
picture, where the same rate emerges as
\begin{eqnarray}\unitlength8mm\label{sigma-+}
W=\begin{picture}(3.75,1.3)\put(.2,.2){\oneloop}\thicklines
\put(1.9,.95){\vector(-1,0){.4}}\put(1.5,-.55){\vector(1,0){.4}}\end{picture}
(1+n_\omega)\delta(\omega-\omega_{\vec q})\\ \nonumber
\end{eqnarray}
with now two types of vertices $-$ and $+$ for the time-ordered and
the anti-time ordered part of the square of the amplitude. Together
with the line sense and the $-$ and $+$ marks at the vertices a unique
correspondence is provided between the oriented
$\stackrel{+~-}{\longrightarrow}$ and
$\stackrel{-~+}{\longrightarrow}$ propagator lines and the initial and
final states. Thus such propagator lines define the densities of
occupied states or those of available states, respectively. Therefore
{\em all standard diagrammatic rules} can be used again. One simply
has to extend those rules to the two types of vertices with marks $-$ and $+$
and the corresponding 4 propagators, the usual time-ordered propagator
$\stackrel{-~-}{\longrightarrow}$ between two $-$ vertices, the
anti-time-ordered one $\stackrel{+~+}{\longrightarrow}$ between two
$+$ vertices and the mixed $\stackrel{+~-}{\longrightarrow}$ or
$\stackrel{-~+}{\longrightarrow}$ ones as densities of occupied and
available states. For details I refer to the textbook of Lifshitz and
Pitaevski\cite{LP}. The advantage of the formulation in terms of
``correlation'' diagrams, which no longer refer to amplitudes but
directly relate to physical observables, like rates, is that now one
is no longer restricted to the concept of asymptotic states. Rather
all internal lines, also the ones which originally correspond to the
``in'' or ``out'' states are now treated on equal footing. Therefore
now one can deal with ``states'' which have a broad mass spectrum and
which therefore appropriately account for the damping of the
particles.  The corresponding Wigner densities
$\stackrel{+~-}{\longrightarrow}$ or $\stackrel{-~+}{\longrightarrow}$
are then no longer on-shell $\delta$-functions in energy (on-mass
shell) but rather acquire a width in terms of the spectral function,
e.g. for non-relativistic particles
\begin{eqnarray}
 \ii G^{-+}&=&\; \stackrel{-~+}{\longleftarrow}\;\;
=\mp f(p)A(p)\nonumber\\[4mm]
\ii G^{+-}&=&\;   \stackrel{+~-}{\longleftarrow}\;\;=(1\mp f(p))A(p)\\[4mm]
A(p)&=&\frac{\Gamma(p)}
{\left(\epsilon+\mu_F -\epsilon_{\vec
   p}^0-\Re\Sigma^R(p)\right)^2
+(\Gamma(p)/2)^2}.\hspace*{-3mm}\nonumber
\end{eqnarray}
Here $f(p)$ is the phase-space occupation at four-momentum
$p=(\epsilon,{\vec p})$, $A$ is the spectral function with the damping
width $\Gamma(p)$ and in-medium on-shell energy $\epsilon_{\vec
  p}^0-\Re\Sigma^R(p)$ and $\mu$ is the chemical potential. In general
all quantities depend on both, energy $\epsilon$ and momentum ${\vec
  p}$.
\begin{figure}\epsfxsize=7cm
\unitlength4.45mm
\begin{picture}(17,3)
\put(-.8,1.5){\contourxy}
\end{picture}
\caption{Closed real-time contour with two external points $x,y$ on the
  contour.} \label{contour-fig}
\end{figure}
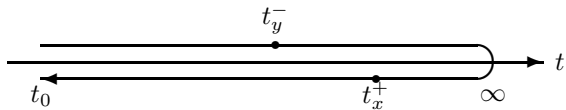
The non-equilibrium theory can entirely be formulated on one special time
contour, the so called closed time path\cite{SKBK}, fig. \ref{contour-fig},
with the time argument running from some initial time $t_0$ to infinity and
back with external points placed on this contour, e.g., for the four different
components of Green's functions or self energies. The special $-+$ or $+-$
components of the self energies define the gain and loss terms in transport
problems, c.f. eq. (\ref{sigma-+}).

\section{Bremsstrahlung from Classical Sources}
For a clarification of the infra-red problem we first discuss two
simple examples of soft modes in hard matter. These are examples in
classical electrodynamics, which both can be solved analytically to a
certain extent: there the hard matter is described either by a
diffusion process or by a random walk problem, respectively \cite{KV}.
As the source particles move non-relativistically these cases do not
suffer from standard pathologies encountered in the hard thermal loop
(HTL) problem of QCD, namely the collinear singularities, where ${\vec
  v}{\vec q}\approx 1$ and from diverging Bose-factors. The advantage
of these examples is that damping can be fully included without
violating current conserving and gauge invariants in the case of
Abelian fields. The closed form results obtained correspond to a
partial re-summation of certain planar diagrams, which just survive in
the classical limit. The problem is related to the
Landau--Pommeranchuk--Migdal effect of Bremsstrahlung in high energy
scattering \cite{LPM}.
\begin{figure}\epsfxsize=7cm
\centerline{\epsfbox{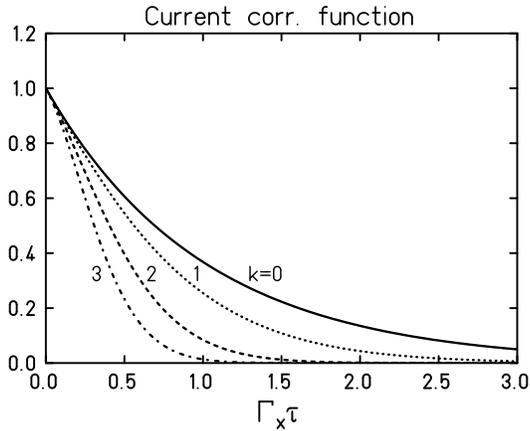}}
\vskip4mm
\caption{Current-current correlation function
in units of $e^2<v^2>$ as a
function of time (in units of $1/\Gamma_x$) for different values of
the photon momentum $q^2=3k^2\Gamma_x^2/{<v^2>}$ with $k=0,1,2,3$.}
\label{CCC}
\end{figure}

The {\em diffusion process} is assumed to be described by a
Fokker-Planck equation for the probability distribution $f$ of
position ${\vec x}$
and velocity ${\vec v}$
\begin{eqnarray}\label{FP}
    &&\frac{\partial}{\partial t} f({\vec x},{\vec v},t)\cr
    &&=
    \left({D\Gamma_x^2}\frac{\partial^2}{\partial {\vec v}^2}
    +\Gamma_x \frac{\partial}{\partial {\vec v}}{\vec v}-{\vec
    v}\frac{\partial}{\partial {\vec x}}\right) f({\vec x},{\vec v},t).
\end{eqnarray}
Likewise fluctuations evolve in time by this equation and this way
determine the correlations.  The two macroscopic parameters are the
spatial diffusion coefficient $D$ and a friction constant $\Gamma_x$
which determines the relaxation rates of velocities (friction due to
collisions with the medium).  In the equilibrium limit
($t\rightarrow\infty$) the distribution attains a Maxwell-Boltzmann
velocity distribution where $T=m\left< {\vec
v}^2\right>/3=mD\Gamma_x$.  The correlation function can be obtained
in closed form and one can discuss the resulting time correlations of
the current at different fixed values of the photon momentum ${\vec
q}$, fig. \ref{CCC} (details are given in ref.\cite{KV}).  For the transverse
part of the correlation tensor this correlation decays exponentially
as $\sim e^{-\Gamma_x\tau}$ at ${\vec q}=0$, and its width further
decreases with increasing momentum $q=|{\vec q}|$. Besides trivial
kinematical factors, the in-medium production rate is given by the
time Fourier transform $\tau\rightarrow\omega$. 
\begin{figure}\epsfxsize=7cm
\centerline{\epsfbox{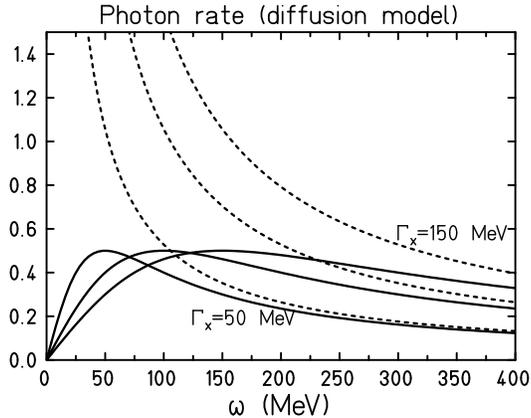}}
\vskip4mm
\caption{Rate of real photons $\di^2
N/(\di\omega\di t)$ in units of $4\pi e^2\left<{\vec v}^2\right>/3$
for a non-relativistic source for $\Gamma_x=$50,100,150 MeV; for
comparison the IQF results (dashed
lines) are also shown.}
\label{Rate}
\end{figure}
Fig. \ref{Rate} displays the corresponding total production rates $\di^2
N/(\di\omega\di t)$ of on-shell photons (number per time and energy; which is
dimensionless) in units of $4\pi e^2\left<{\vec v}^2\right>/3$. One sees that
the hard part of the spectrum behaves as expected, namely, like in the IQF
approximation the rate grows proportional to $\Gamma_x$ and this way
proportional to the microscopic collision rate $\Gamma$ (c.f. below). However
independent of $\Gamma_x$ the rate saturates at a value of $\sim 1/2$ in these
units around $\omega\sim\Gamma_x$, and the soft part shows the inverse
behavior.  That is, with increasing collision rate the production rate is
more and more suppressed! This is in line with the picture that such photons
cannot resolve the individual collisions any more. Since the soft part of the
spectrum behaves like $ \omega/\Gamma_x$, it shows a genuine non-perturbative
feature which cannot be obtained by any power series in $\Gamma_x$.  For
comparison: the dashed lines show the corresponding IQF yields, which agree
with the correct rate for the hard part while they completely fail and diverge
towards the soft end of the spectrum.  For non-relativistic sources
$\left<{\vec v}^2\right>\ll 1$ one can ignore the additional $q$-dependence
(dipole approximation; c.f. fig. \ref{CCC}) and the entire spectrum is
determined by one macroscopic scale, the relaxation rate $\Gamma_x$. This
scale provides a quenching factor
\begin{equation}\label{suppression}
C_0(\omega)=\frac{\omega^2}{\omega^2+\Gamma_x^2}\; .
\end{equation}
by which the IQF results have to be corrected in order to account for
the finite collision time effects in dense matter.\\[2mm] 

In the {\em microscopic Langevin picture} one considers a classical process,
where hard scatterings occur at random with a constant {\em mean collision
  rate} $\Gamma$. These scatterings consecutively change the velocity of a
point charge from ${\vec v}_m$ to ${\vec v}_{m+1}$ to ${\vec v}_{m+2}$,
$\dots$ (in the following subscripts $m$ and $n$ refer to the collision
sequence). In between scatterings the charge moves freely. For such a multiple
collision process some explicit results can be given, since the correlated
probability to find the charge at time $t_1$ and $t_2$ at two different
segments with $n$ scatterings in between follows from the iterative folding of
the exponential decay law with decay time $1/\Gamma$. Therefore the space
integrated current-current correlation function takes a simple Poisson form
\begin{eqnarray}\label{Apoisson}
\ii\Pi^{\mu\nu}_{-+}   &\propto&\int \di^3x_1\di^3x_2
\makebox[3.5cm][l]{$ \left<j^{\mu}({\vec
      x_1},t-\mbox{$\frac{\tau}{2}$})j^{\nu}({\vec 
 x_2},t+\mbox{$\frac{\tau}{2}$})\right>$}\cr
&=& e^2 
   \left<v^{\mu}(0)v^{\nu}(\tau)\right>\cr\cr
&=&
   e^2e^{-|\Gamma\tau|}\sum_{n=0}^\infty
   \frac{|\Gamma\tau|^n}{n!} \left< v^{\mu}_m v^{\nu}_{m+n}\right>_m
\end{eqnarray} 
with $v=(1,{\vec v})$.  This result represents a genuine multiple collision
description of the correlation function.  Here $\left<\dots\right>_m$ denotes
the average over the discrete collision sequence $\{m\}$. This form, which one
writes down intuitively, directly includes what one calls {\em damping} in the
corresponding quantum case. Fourier transformed it determines the spectrum in
completely regular terms (void of any infra-red singularities) where each term
describes the interference of the photon being emitted at a certain time or
$n$ collisions later.

In special cases where velocity fluctuations are degraded by a
constant fraction $\alpha$ in each collision, such that $\left< {\vec
v}_m\cdot{\vec v}_{m+n} \right>_m= \alpha^n\left< {\vec v}_m\cdot{\vec
v}_{m} \right>_m$, one can re-sum the whole series in (\ref{Apoisson})
and thus recover the relaxation result with $2\Gamma_x\left< {\vec
v}^2 \right>= \Gamma\left< ({\vec v}_m-{\vec v}_{m+1})^2 \right>$ at
least for ${\vec q}=0$ and the corresponding quenching factor
(\ref{suppression}).  

This clarifies that the diffusion result represents a re-summation of
the Langevin multiple collision picture and altogether only
macroscopic scales are relevant for the form of the spectrum and not
the details of the microscopic collisions. Note also that the
classical results, both for the diffusion equation (c.f. fig. \ref{CCC}) and
for the Langevin process fulfill the classical version
($\hbar\rightarrow 0$) of the sum rules discussed in refs. \cite{DGK,KV}.
\section{Radiation on the Quantum level}
We have seen that on the classical level the problem of radiation from
dense matter can be solved quite naturally and completely at least for
simple examples, and figs. \ref{CCC} and \ref{Rate} display the main
physics. They show, that the {\em damping} of the particles due to
scattering is an important feature, which in particular has to be
included right from the onset.  This does not only assure results
which no longer diverge, but also provides a systematic and convergent
scheme. On the {\em quantum level} such problems requires techniques beyond the
standard repertoire of perturbation theory or the quasi-particle
approximation.

\noindent\parbox[t]{4.5cm}{The production or
absorption rates are given by photon self energy diagrams of the type to
the\hfill right\hfill with\hfill an\hfill in--\hfill and} 
\parbox[t]{2.6cm}{\unitlength7mm \begin{picture}(5,.7)
   \put(0.4,-.6){\hatchedself} 
\end{picture}}\\[.5ex]
outgoing photon line (dashed). The hatched loop area denotes all strong
interactions of the source. The latter give rise to a whole series of
diagrams.  As mentioned, for the particles of the source, e.g. the
nucleons, one has to re-sum Dyson's equation with the corresponding
full complex self energy in order to determine the full Green's
functions in dense matter. Once one has these Green's functions
together with the interaction vertices at hand one could in principle
calculate the required diagrams. However both, the computational
effort to calculate a single diagram and the number of diagrams, are
increasing dramatically with the loop order of the diagrams, such that
in practice only lowest order loop diagrams can be considered in the
full quantum case. In certain limits some diagrams drop out. We could
show that in the {\em classical limit} of the quantum description,
which in this case implies the hierarchy $\omega,|{\vec q}|,\Gamma\ll
T\ll m$ together with low phase-space occupations for the source, i.e.
$f(x,p)\ll 1$, only the following set of diagrams survive
\begin{equation}\label{classicaldiagram}
\unitlength6mm
\begin{picture}(12,4)\put(0,.2){\classdiagram}\end{picture}
\end{equation}
In these ``Langevin'' diagrams the bold lines denote the
full nucleon Green's functions which also include the damping width, the black
blocks represent the effective nucleon-nucleon interaction in matter, and the
full dots the coupling vertex to the photon. Each of these diagrams with $n$
interaction loop insertions just corresponds to the $n^{th}$ term in the
classical Langevin result (\ref{Apoisson}). Thus the classical multiple
collision example provides a quite intuitive picture about such diagrams.
Thereby the diagram of order $n$ describes the interference of the amplitude
where the photon is ''emitted'' at some time and that where it is ''emitted''
$n$ collisions later. Further details are given in\cite{KV}.

\section{$\Phi$-derivable approximations}

Following Luttinger, Ward \cite{Luttinger}, and the reformulation by
Cornwall, Jackiw and Tomboulis \cite{Cornwall} using path-integral
methods for equilibrium case, the generating functional
$\Gamma\{\phi,\Gr\}$ for the equations of motions, both, for the classical
fields $\phi=<\medhat{\phi}>$ and Dyson's equation for the
propagators $\Gr$, can be expressed in terms of an auxiliary functional
$\Phi$, where $\Phi$ is solely given in terms of full, i.e. re-summed,
propagators $\Gr$ and full classical fields $\phi$. Following \cite{IKV1}
it can be generalized to the real time case with the diagrammatic
representation \unitlength=.8cm
%
\begin{eqnarray}\label{keediag}
&&\ii\Gamma\left\{\phi,\Gr\right\} = \ii
\Gamma^0\left\{\Gr^0\right\}  +\oint \di x \Lg^0\{\phi,\partial_\mu\phi\}
\nonumber 
\\
&&
+
\underbrace{\sum_{n_\Se}\vhight{1.6}\frac{1}{n_\Se}\GlnG0Sa}
_{\displaystyle \pm \ln\left(1-\odot\Gr^{0}\odot\Se\right)}
\underbrace{-\vhight{1.6}\GGaSa}
_{\displaystyle \pm \odot\Gr\odot\Se\vphantom{\left(\Ga^{0}\right)}}
\nonumber\\
&&\underbrace{+\vhight{1.6}\sum_{n_\lambda}\frac{1}{n_\lambda}
\Dclosed{c2}{\thicklines}}
_{\displaystyle\vphantom{\left(\Ga^{0}\right)} 
+\ii\Phi\left\{\phi,\Gr\right\}}.
\end{eqnarray}
%
Here upper signs relate to fermion quantities, whereas lower signs, to
boson quantities. Thereby $n_\Se$ counts the number of self-energy
$\Se$-insertions in the ring diagrams, while for the closed diagram of
$\Phi$ the value $n_\lambda$ counts the number of vertices building up
the functional $\Phi$.  Due to this factor such a set of diagrams is
not resumable in the standard diagrammatic sense. The $\Gamma^0$
solely depends on the unperturbed propagator $\Gr^0$ (thin line) and,
hence, is treated as a constant with respect to the functional
variations in $\Gr(x,y)$ and $\phi(x)$.  The diagrams contributing to $\Phi$
are given in terms of full propagators $\Gr$ (thick lines) and
classical fields $\phi$.  As a consequence, these $\Phi$-diagrams have
to be {\em two-particle irreducible} (label $c2$), i.e. they cannot be
decomposed into two pieces by cutting two propagator lines.  The
latter property matches diagrammatic rules for the re-summed
self-energy $\Se(x,y)$ and the source current $J(x)$ of the classical
field equations, which results from functional variation of $\Phi$
with respect to any propagator $\Gr(y,x)$, i.e.
\begin{eqnarray}\label{varphi}
-\ii \Se =\mp \delta \ii \Phi / \delta \ii \Gr,\quad
\ii J = \delta \ii \Phi / \delta \phi  . 
\end{eqnarray}
It directly follows from the stationarity condition of $\Gamma$
(\ref{keediag}) with
respect to variations of $\Gr$ and $\phi$ on the contour 
%
\begin{eqnarray}
\label{varG/phi}
\delta \Gamma \left\{\phi,\Gr \right\}/ \delta \Gr = 0,\quad
\delta \Gamma \left\{\phi,\Gr \right\}/ \delta \phi = 0,
\end{eqnarray}
%
which indeed provides the Dyson equation with self-energy consistent
with respect to the $\Phi$-functional and the classical field
equation.  In graphical terms, the variation (\ref{varphi}) with
respect to $\Gr$ is realized by opening a propagator line in all
diagrams of $\Phi$.  The resulting set of thus opened diagrams must
then be that of proper skeleton diagrams of $\Se$ in terms of {\em
  full propagators}, i.e.  void of any self-energy insertions.

In order to arrive at a closed and consistent scheme we consider the
so-called $\Phi$-derivable approximation, first introduced by Baym
\cite{Baym} based on ref. \cite{KaBaym} within linear response to
external perturbation of equilibrated systems. They used the
corresponding imaginary time formulation.  A $\Phi$-derivable
approximation is constructed by confining the infinite set of diagrams
for $\Phi$ to either only a few of them or some sub-series of them.
Note that $\Phi$ itself is constructed in terms of ``full'' Green's
functions and classical fields, where ``full'' now takes the sense of
solving self-consistently the Dyson and Classical field equation with
the driving terms $\Se$ and $J$ derived from this $\Phi$ through
relation (\ref{varphi}). It means that even restricting ourselves to a
single diagram in $\Phi$, in fact, we deal with a whole sub-series of
perturbation theory diagrams, and ``full'' takes the sense of the sum
of this whole sub-series.  Thus, a $\Phi$-derivable approximation
offers a natural way of introducing closed, i.e.  consistent
approximation schemes based on summation of diagrammatic sub-series.
In order to preserve the original symmetry of the exact $\Phi$ we
postulate that the set of diagrams defining the $\Phi$-derivable
approximation complies with all such symmetries. As a consequence,
approximate forms of $\Phi^{\scr{(appr.)}}$ define {\em effective}
theories, where $\Phi^{\scr{(appr.)}}$ serves as a generating
functional for approximate self-energies $\Sa^{\scr{(appr.)}}(x,y)$
and source currents $J(x)$ through relation (\ref{varphi}), which then
enter as driving terms for the Dyson equations. The propagators
solving this set of Dyson equations are still called ``full'' in the
sense of the $\Phi^{\scr{(appr.)}}$-derivable scheme.  Below, we omit
the superscript ``appr.''.

\section{Generalized Kinetic Equation}\label{sect-Kin-EqT}  
\subsection{Gradient Expansion Scheme}

For slightly inhomogeneous and slowly evolving systems, the degrees of
freedom can be subdivided into rapid and slow ones. Any kinetic
approximation is essentially based on this assumption.  Then for any
two-point function $F(x,y)$, one separates the variable $\xi
=(t_1-t_2, \vec{r_1}-\vec{r_2})$, which relates to rapid and
short-ranged microscopic processes, and the variable $X=
\frac{1}{2}(t_1+t_2,\vec{r_1}+\vec{r_2})$, which refers to slow and
long-ranged collective motions. The Wigner transformation, i.e.  the
Fourier transformation in four-space difference $\xi=x-y$ to
four-momentum $p$ of the contour decomposed components of
$F^{ij}$,$i,j\in\{-+\}$ 
%
\begin{equation}
\label{W-transf} 
F^{ij}(X;p)=\int \di \xi e^{\ii p\xi}
F^{ij}\left(X+\xi/2,X-\xi/2\right)
\end{equation}
%
leads to a (co-variant) four phase-space formulation of two-point functions.
The Wigner transformation of Dyson's equation (\ref{varG/phi}) in $\{-+\}$
notation is straight forward. For details and the extensions to include the
coupling to classical field equations we refer to ref. \cite{IKV1}.

Standard transport descriptions usually involve two approximation steps: (i)
the gradient expansion for the slow degrees of freedom, as well as (ii) the
quasi-particle approximation for rapid ones. We intend to avoid the latter
approximation and will solely deal with the gradient approximation for slow
collective motions by performing the gradient expansion of the coupled Dyson
equations. This step indeed preserves all the invariances of the $\Phi$
functional in a $\Phi$-derivable approximation.
\subsection{Generalized Kinetic Equation in Physical Notation}

It is helpful to avoid all the imaginary factors inherent in the
standard Green's function formulation and change to quantities which
are real and in the homogeneous limit positive and therefore have a
straight physical interpretation much like for the Boltzmann equation.
We define
%
\begin{eqnarray}
\label{F}
\Fd (X,p) &=& \A (X,p) \fd (X,p)
 = \ii (\mp )\Gr^{-+} (X,p) , \nonumber\\
\Fdt (X,p) &=& \A (X,p) [1 \mp \fd (X,p)] = \ii \Gr^{+-} (X,p)
\end{eqnarray}
%
for the generalized Wigner functions $\F$ and $\Ft$ and the
corresponding {\em four} phase space distribution functions $\fd(X,p)$
and Fermi/Bose factors $[1 \mp \fd (X,p)]$. Here
%
\begin{eqnarray}
\label{A}
 A (X,p) \equiv -2\Im \Gr^R (X,p) = \Fdt \pm \Fd
\end{eqnarray}
is the spectral function. According to retarded relations between Green's
functions $\Gr^{ij}$, {\em only two of these real functions are required for a
  complete description of the system's evolution}.

The reduced gain and loss rates and total width of the collision
integral are
%
\begin{eqnarray}
\label{gain}
\Ldt (X,p) &=&  \ii (\mp ) \Se^{-+} (X,p),\nonumber\\
\Ld (X,p)  &=&  \ii \Se^{+-} (X,p) . 
\end{eqnarray}
%
They determine the damping width
%
\begin{eqnarray}
\label{G-def}
\Gamma (X,p)&\equiv& -2\Im \Se^R (X,p)\cr &=& \Ld (X,p)\pm\Ldt (X,p), 
\end{eqnarray}
%
where $\Gr^R$ and $\Se^R$ are the retarded propagator and self-energy,
respectively. The opposite combinations
%
\begin{eqnarray}
\label{Fluc-def}
I (X,p) = \Ldt (X,p)\mp\Ld (X,p), 
\end{eqnarray}
%
determines the fluctuations.
 
In terms of the new notation (\ref{F}) - (\ref{G-def}) and in the first
gradient approximation the {\em generalized kinetic} equation for $F$
takes the form
%
\begin{equation}
\label{keqk}
\Do 
\F (X,p) - 
B
= C (X,p) 
\end{equation}
%
with the differential drift operator (for simplicity in
non-relativistic kinematics)
%
\begin{eqnarray}\label{Drift-O}
\Do = 
\left(
\vu_{\mu} - 
\frac{\partial \Re\Sa^R}{\partial p^{\mu}} 
\right) 
\partial^{\mu}_X + 
\frac{\partial \Re\Sa^R}{\partial X^{\mu}}  
\frac{\partial }{\partial p_{\mu}} 
\end{eqnarray}
with $v^\mu=(1,\vec{p}/m)$. Further $C(X,p)$ and $B(X,p)$ are the collision
and a fluctuation term, respectively
%
\begin{eqnarray}
\label{Coll(kin)}
C (X,p) &=& 
\Ldt (X,p) \Ft (X,p) 
- \Ld (X,p) \F (X,p) \nonumber\\
B&=&\Pbr{\Ldt , \Re\Gr^R} 
.  
\end{eqnarray}
%

We need still one more equation, which can be provided by the retarded Dyson
equation. In terms of the new notation it takes the simple form
%
\begin{eqnarray}
\label{keqX}
\Do \Ga^R (X,p) + \frac{\ii}{2} \Pbr{\Gm , \Ga^R} &=&0, \\ 
\label{meqX}
\left( M (X,p)+\frac{\ii}{2} \Gm (X,p) \right)\Ga^R (X,p) &=& 1 , 
\end{eqnarray}
%
with the  ''mass'' function 
%
\begin{eqnarray}\label{meqx}\label{M}
M(X,p)=p_0 -\frac{1}{2m}\vec{p}^2 -\Re\Se^R (X,p), 
\end{eqnarray}
which relates to the drift operator via $Df=\Pbr{M,f}$ for any four
phase-space function $f$.  Subset (\ref{keqX}) - (\ref{meqX}) is
solved by \cite{KBBM}
%
\begin{eqnarray}
\label{Asol}\label{Xsol}
&&\Gr^R=\frac{1}{M(X,p)+\ii\Gamma(X,p)/2}\\ \nonumber
&&\Rightarrow
\left\{\begin{array}{rcl}
A (X,p) &=&\displaystyle
\frac{\Gamma (X,p)}{M^2 (X,p) + \Gamma^2 (X,p) /4},\\[4mm]
\Re\Gr^R (X,p) &=& \displaystyle 
\frac{M (X,p)}{M^2 (X,p) + \Gamma^2 (X,p) /4}.
\end{array}\right.\hspace*{-0.2cm} 
\end{eqnarray}
%
The canonical equal-time (anti) commutation relations for (fermionic) bosonic
field operators imply sum-rules for the spectral function which for
non-relaticistic kinematics read
\begin{eqnarray}
\label{A-sumf} 
\int_{-\infty}^{\infty} \frac{\di p_0}{2\pi} 
A(X,p) &=& 1 .
\end{eqnarray}
%

We now provide a physical interpretation of various terms in the generalized
kinetic equation (\ref{keqk}).  The drift term $\Do \Fd$ on the l.h.s. of eq.
(\ref{keqk}) is the usual kinetic drift term including the corrections from
the self-consistent field $\Re\Se^R$ into the convective transfer of real and
also virtual particles.  In the collision-less case $C=B=0$, i.e. \mbox{$\Do
  \Fd=0$} (Vlasov equation), the quasi-linear first order differential
operator $\Do$ defines characteristic curves. They are the standard classical
paths in the Vlasov case. Thereby the four-phase-space probability $\Fd(X,p)$
is conserved along these paths. The formulation in terms of a Poisson bracket
in four dimensions implies a generalized Liouville theorem. In the collisional
case both, the collision term $C$ and the fluctuation term $B$ change the
phase-space probabilities of the ''generalized'' particles during their
propagation along the ''generalized'' classical paths given by $\Do$. We use
the term ''generalized'' in order to emphasize that particles are no longer
bound to their mass-shell, $M=0$, during propagation due to the collision
term, i.e.  due decay, creation or scattering processes.

The r.h.s. of eq. (\ref{keqk}) specifies the collision term $C$ in terms of
gain and loss terms, which also can account for multi-particle processes.
Since $\Fd$ includes a factor $A$, $C$ further deviates from the standard
Boltzmann-type form in as much that it is multiplied by the spectral function
$A$, which accounts for the finite width of the particles.

The additional Poisson-bracket term
%
\begin{eqnarray}
\label{backflow}  
B&=&\Pbr{\Ldt,\Re\Gr^R}=\frac{M^2-\Gamma^2/4}{(M^2+\Gamma^2/4)^2}\;
   \Do\;\Ldt\cr
   &&+\frac{M\Gamma}{(M^2+\Gamma^2/4)^2}\Pbr{\Ldt,\Gamma}
\end{eqnarray}
%
is special. It contains genuine contributions from the finite mass width of
the particles and describes the response of the surrounding matter due to
fluctuations. This can be seen from the conservation laws discussed below. In
particular the first term in (\ref{backflow}) gives rise to a back-flow
component of the surrounding matter. It restores the Noether currents as the
conserved ones from the intuitively expected sum of convective currents
arising from the convective $\Do\Ft$ terms in (\ref{keqk}).  The second term
of (\ref{backflow}) gives no contribution in the quasi-particle limit of small
damping width limit and represents a specific off mass-shell response, c.f.
\cite{MLipS,IKV2}.

\subsection{Conservations of the Current and Energy--Momentum}
\label{Conservation-L}

Special combinations of the transport equations (\ref{keqk}) and the
corresponding one for $\Ft$
weighted with $e$ and $p^\nu$, and integrated over momentum give rise to the
charge and energy--momentum conservation laws, respectively, with the Noether
charge current and Noether energy--momentum tensor defined by the following
expressions
%
\begin{eqnarray}
\label{c-new-currentk}\nonumber 
j^{\mu} (X) 
&=&\frac{e}{2}\mbox{Tr} \int \dpi{p}
\vu^{\mu} 
\left(\Fd (X,p) \mp \Fdt (X,p) \right),\hspace*{-1cm} \\
\label{E-M-new-tensork}
\Theta^{\mu\nu}(X)
&=&\frac{1}{2}\mbox{Tr} \int \dpi{p} 
\vu^{\mu} p^{\nu} 
\left(\Fd (X,p) \mp \Fdt (X,p) \right)\hspace*{-1.2cm}\cr
&&+ g^{\mu\nu}\left(
{\cal E}^{\scr{int}}(X)-{\cal E}^{\scr{pot}}(X)
\right).  
\end{eqnarray}
%
Here 
%
\begin{eqnarray}
\label{eps-int} 
{\cal E}^{\scr{int}}(X)=\left<-\Lint(X)\right>
=\left.\frac{\delta\Phi}{\delta\lambda(x)}\right|_{\lambda=1}
\end{eqnarray}
%
is the interaction energy density, which in terms of $\Phi$ is given
by a functional variation with respect to a space-time dependent coupling
strength of $\Lint\rightarrow\lambda(x)\Lint$, c.f. ref.  \cite{IKV1}. The
potential energy density ${\cal E}^{\scr{pot}}$ takes the form
%
\begin{eqnarray}
\label{eps-potk}
{\cal E}^{\scr{pot}}
= \frac{1}{2}\mbox{Tr}
\int\dpi{p} \left[
\Re\Sa^R \left(\Fd\mp\Fdt\right)
+ \Re\Ga^R I\right]\nonumber
\end{eqnarray}
%
where $I=\Gb\mp\Gbt$.  Whereas the first term complies with
quasi-particle expectations, namely mean potential times density, the
second term displays the role of fluctuations $I=\Gb\mp\Gbt$ in the
potential energy density. Since in many cases interaction and
potential energy are proportional to each other, the same statement
applies to the interaction energy, too. This fluctuation term
precisely arises form the $B$-term in the kinetic eq. (\ref{keqk}),
discussed around eq. (\ref{backflow}). It restores that the Noether
expressions (\ref{E-M-new-tensork}) are indeed the conserved
quantities. In this compensation we see the essential role of the
fluctuation term in the generalized kinetic equation. Dropping or
approximating this term would spoil the conservation laws. Indeed,
both expressions in (\ref{E-M-new-tensork}) comply exactly with the
generalized kinetic equation (\ref{keqk}), i.e. they are exact
integrals of the generalized kinetic equations of motion. As usual the
existence of such conservation laws require certain invariances which
lead to certain consistency relations.  In ref. \cite{IKV1,IKV2} it
has been shown that these are met if all the self-energies are
$\Phi$-derivable.
  
In the field theoretical case there are contributions in
(\ref{E-M-new-tensork}), describing modifications of the vacuum-polarization
in matter.  These terms are generally ultra-violet divergent, and hence, have
to be properly renormalized on the vacuum level. Alongside the
spectral sum-rule (\ref{A-sumf}) gets modified by wave-function
renormalization.

\subsection{Multiprocess Decomposition of $\Phi$-Derivable
Collision Term} 
\label{Multiprocess-Phi}

To be specific we consider a system of fermions interacting via a two-body
potential $V=V_0 \delta(x-y)$, and, for the sake of simplicity, disregard its
spin structure, by reducing spin and anti-symmetrization effects to a
degeneracy factor $d$.  To derive the decomposition of a $\Phi$-derivable
collision term, we employ the same rules as described in ref. \cite{IKV2}.

In the first example we consider the generating functional $\Phi$ to be
approximated by the following two diagrams \vspace*{-6mm}
%
\begin{eqnarray}
\label{Phi-12} 
\ii \Phi =  \frac{1}{2} \loopa + \frac{1}{4} \loopb{c} ,
\\ 
\nonumber
\vspace*{16mm} 
\end{eqnarray}
%
the dashed line illustrating the decomposition.
In the $\{-+\}$ matrix notation of the Green's functions one can easily
see that one-point diagrams do not contribute to the collision term, while
decomposing the second one along the dashed line leads to a purely local
result
%
\begin{eqnarray}
\label{C2} 
&&C_{(2)} 
= d^2
\int \dpi{p_1} \dpi{p_2} \dpi{p_3}
\left|\;\; \wa{-} \;\; \right|^2
\\
\nonumber 
&& \times 
\delta^4\left(p + p_1 - p_2 - p_3\right) 
\left(
\F_2\F_3 \Ft\Ft_1 -
\Ft_2\Ft_3 \F\F_1
\right), 
\end{eqnarray}
%
where the brief notation $\F_i = \F(X,p_i)$ etc.  is used. This collision
integral has precisely the form of the binary collision term of
Boltzmann--Uehling--Uhlenbeck (BUU), except for the fact that the distribution
functions are not constrained by the mass shell. The binary transition rate
%
\begin{eqnarray}
\label{R2} 
R_2^{(2)} = V_0^2 = \left|\;\; \wa{-} \;\; \right|^2
\end{eqnarray}
%
is non-negative in this case. 

The picture becomes more complicated, if $\Phi$ involves diagrams of
higher orders. For instance, let us add the following three point diagram
to $\Phi$, which is next in a series of ring diagrams, i.e.
%
\begin{eqnarray}
\label{Phi-123} 
&&\ii \Phi = \ii\left(\Phi_{(1)} + \Phi_{(2)} + \Phi_{(3)} \right)  
\nonumber 
\\
&&=   \loopa + \frac{1}{2} \loopb{c} 
+ \frac{1}{3} \loopc{c}
\\ 
\nonumber
\vspace*{18mm} 
\end{eqnarray}
%
where one possible decomposition is illustrated by the dashed line. The
corresponding self-energy becomes
%
\begin{eqnarray}
\label{Pi-123} 
&&-\ii \Se = -\ii\left(\Se_{(1)} + \Se_{(2)} + \Se_{(3)} \right) = 
\nonumber 
\\
&&  \pia \; + \;\;\pib \; + \;\;\pic  
\end{eqnarray}
%
Now the collision term contains a {\em non-local} part due to the last
diagram. The local part can easily be derived in the form
%
\begin{eqnarray}
\label{C30} 
&&C^{\scr{loc}}_{(2)} + C^{\scr{loc}}_{(3)} 
= d^2
\int \dpi{p_1} \dpi{p_2} \dpi{p_3}
\nonumber 
\\[3mm]
&&\times \left(
\left|\;\; \wa{-} \;\;+ \!\!\!\!\!\wb{-} \;\;\right|^2 - 
\left|\!\!\!\!\!\wb- \;\;\right|^2\right)
\\[1mm]
&&\times 
\delta^4\left(p + p_1 - p_2 - p_3\right) 
\left(
\F_2\F_3 \Ft\Ft_1 -
\Ft_2\Ft_3 \F\F_1
\right),  
\nonumber 
\end{eqnarray}
%
where all the vertices in the off-shell scattering amplitudes are of the same
sign, say $"-"$ for definiteness, i.e., there are no $"+-"$ and $"-+"$ Green's
functions left. The quantity $C^{\scr{loc}}_{(2)}+C^{\scr{loc}}_{(3)}$ is
again of the Boltzmann form
%
\begin{eqnarray}
\label{R3}\nonumber 
R^{(2)}_2 +R^{(3)}_2  = 
\left|\;\; \wa{-} \;\; + \!\!\!\!\!\wb{-} \;\;\right|^2 - 
\left|\!\!\!\!\!\wb{-} \;\;\right|^2,
\end{eqnarray}
%
the sub-label $2$ denoting that $2$ pairs of particle-hole lines are affected
by the decomposition cut. It can be shown that under normal circumstances also
this rate coefficient is positive. 
\subsection{Kinetic Entropy}
Ignoring higher order gradients the generalized kinetic equation (\ref{keqk})
provides us with the following relation
%
\begin{eqnarray}
\label{s-eq.} 
\partial_\mu s^\mu (x) =\sum_a
\int \dpi{p} \ln \frac{\Ft_a}{\F_a} C_a (x,p),   
\end{eqnarray}
%
where the quantity 
%
\begin{eqnarray}
\label{S(gen)} 
&&s^\mu  = \sum_a s^\mu_a =
\sum_a\int \dpi{p}\cr
&&\left[
\left(
\vu^{\mu} - \frac{\partial \Re\Se^R_a}{\partial p_{\mu}}
\right)\left(
\mp \Ft_a \ln \frac{\Ft_a}{A_a} - \F_a \ln \frac{\F_a}{A_a} 
\right)
\right.
\nonumber 
\\
&&- 
\left. 
\Re\Gr^R_a
\left(
\mp \frac{\partial \Get^a}{\partial p_{\mu}} \ln \frac{\Ft_a}{A_a} 
- 
\frac{\partial \Ge^a}{\partial p_{\mu}} \ln \frac{\F_a}{A_a} 
\right)
\right], 
\end{eqnarray}
%
obtained from the l.h.s. of the kinetic equation is interpreted as the
Markovian part of the entropy flow. Here we have introduced a
summation over $a$ denoting the different particle species and
intrinsic quantum numbers for a multi-component system.  The
interesting aspect is that for special local collision terms $C_a$ as
the ones discussed above the r.h.s. of (\ref{s-eq.}) can be shown to
be non-negative and therefore gives rise to an H-theorem. Again the
functional properties of $\Phi$ have been used. The positivity of
r.h.s. of (\ref{s-eq.}  is exactly given for $\Phi$-functionals with
two internal points for which in the equilibrium limit the zero
component of the non-equilibrium entropy flow (\ref{S(gen)}) agrees
with the corresponding equilibrium entropy. Memory corrections as
contained in $\Phi$-functionals with more than two points give rise to
extra gradient terms which contribute to the entropy flow. For details
we refer to our forthcoming paper \cite{IKV2}.

\section{Conclusion}
In the first part of this talk the problem of soft modes in hard dense
matter is discussed under circumstances which can be treated
completely in analytical terms. The hard modes are described by a
Fokker-Planck equation. They couple to a classical Maxwell field for
the soft modes, c.f. Fig.  \ref{soft-modes-fig}.  For this Abelian
case the result is conserving and completely gauge invariant even
though the damping of the source particles is fully included. The
friction coefficient (closely related to the damping) determines the
scale that separates soft from hard modes. This classical scheme is
seen to re-sum a certain set of planar diagrams in the quantum case,
which do survive in the classical limit. Such concepts are quite a
general.  In recent times they have been applied to the hard thermal
loop (HTL) re-summation \cite{BraatenPisarski} in terms of classical
transport \cite{Blaizot,Jackiw93}. In the non-abelian QCD case, however,
in order to preserve gauge invariance, the transport part is limited
to the collision-less Vlasov equation, i.e. neglecting the damping of
the source particles.  A historical hard-loop re-summation scheme is
the Fermi-liquid problem, where soft RPA-modes are treated by the
coupling to the fermions in the Fermi-sea, the latter representing the
hard modes, c.f. Fig.  \ref{soft-modes-fig}.
\def\Blue{}\def\Black{}\def\Red{}\def\Green{}
\begin{figure}
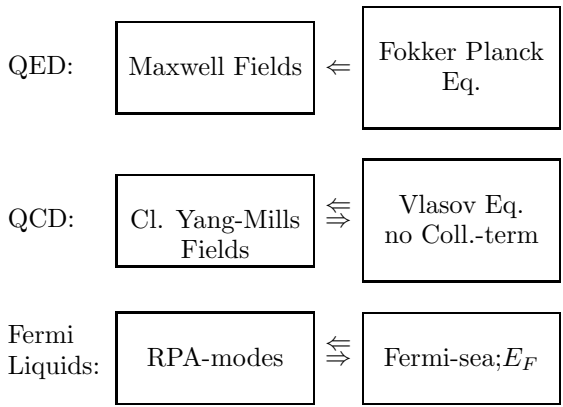

\normalsize\Red
\hspace*{-0.5cm}
\begin{tabular}{lccc}
\Black QED:&\fbox{\parbox{2.4cm}{\center Maxwell Fields\\ $ $}}&
\Black$\Leftarrow$&
{\Blue\fbox{\parbox{2.4cm}{\center Fokker Planck Eq.\\ $ $}}}\\ $ $\\
\Black QCD:&\fbox{\parbox{2.4cm}{\center Cl. Yang-Mills Fields$ $}}&
$\Black\stackrel{\mbox{$\Leftarrow$}}{\Rightarrow}$&
{\Blue\fbox{\parbox{2.4cm}{\center Vlasov Eq.\\ no Coll.-term\\ $
      $}}}\\ $ $\\
\parbox{1.3cm}{$ $\\ \Black Fermi\\
\Black Liquids:\\}&\fbox{\parbox{2.4cm}{\center RPA-modes\\ $ $}}&
$\Black\stackrel{\mbox{$\Leftarrow$}}{\Rightarrow}$&
{\Blue\fbox{\parbox{2.4cm}{\center Fermi-sea;$ E_F$\\ $ $}}}
\end{tabular}\\[0.6cm]
\caption{Hard Loop Re-Summation}\label{soft-modes-fig}
\end{figure}

In practical terms we have seen that the spectrum of soft particles
resulting from collisions in dense matter can no longer appropriately
be described by the {\em quasi-particle} approximation, since it leads
to divergent results in the soft limit. Rather the finite time between
successive collisions and the ensuing relaxation rates $\Gamma_x$ in
dense matter lead to a considerable quenching of the production rate,
e.g. at small photon energies.  This can be compiled in the simple
quenching factor (\ref{suppression}). Fig. \ref{Rate} summarizes the
main behavior, which also is relevant in a quantum treatment of the
source.

In the second part a scheme is presented that leads to self-consistent
conserving transport equations. There we essentially followed ideas
suggested by Kadanoff and Baym in particular. The central quantity is a
functional $\Phi$ which generates the driving terms for the classical
field and transport equations. It can be truncated at any desired loop
order and still provides equations which fulfill conservation
laws. We explicitly constructed the energy momentum tensor for this
$\Phi$-derivable method. The gradient approximation provided equations
of classical type for the phase-space distribution functions in four
dimensions. At no place the quasi-particle approximation was
necessary. Alongside from the $\Phi$-derivable properties a kinetic
entropy could be derived, which in some cases leads to an H-theorem.

In summary the method has the following advantages:
\itemsep-1mm
\begin{itemize}\itemsep-1mm
\item[{\Green \boldmath{$\clubsuit$}}] provides a self-consistent \& conserving
  transport scheme;
\item[{\Green \boldmath{$\clubsuit$}}] allows to include classical fields
  (soft modes);
\item[{\Green \boldmath{$\clubsuit$}}] includes all QM effects that are
  accounted for in the corresponding equilibrium treatment;
\item[{\Green \boldmath{$\clubsuit$}}] has no limitation to small widths;
\item[{\Green \boldmath{$\clubsuit$}}] includes delay-time, drag \&
  back flow, and memory effects.
\end{itemize}
There are two limitations: first, the derivation is limited to slow
space-time variations of the macroscopic quantities; secondly,
local symmetries, like gauge invariance, may be violated by such
re-summation schemes. The latter problem is inherent to all approaches,
based on truncated self-consistent dynamical equations.

Our considerations are of particular importance for the theoretical
description of nucleus-nucleus collisions at intermediate to
relativistic energies. The kinematical feature are such that damping
effects play an essential role, i.e. the energy uncertainty of
the particles is comparable with their mean kinetic energy! In
particular the bulk production and absorption rates of all particles
with masses less than $T$, if calculated in standard quasi-particle
approximation, are seriously subjected to the here discussed effects.

In summary, the combined effort from many sides to include the finite
width of the particles in dense matter, may give hope for a unified
transport theory which appropriately describes both, the propagation
of resonances and of off-shell particles in the dense matter
environment.
\topsep=0mm


\begin{thebibliography}{99}
\bibitem{SKBK} J. Schwinger, {J. Math. Phys,} {\bf 2} (1961) 407;
L. P. Kadanoff and G. Baym, Quantum Statistical Mechanics (Benjamin,
1962); L. M. Keldysh, {ZhETF} {\bf 47} (1964) 1515; in
Engl. translation Sov. Phys. JETP{\bf 20 } (1965) 1018.
\bibitem{BB}B.Bezzerides and D.F. DuBois, Ann. Phys. (N.Y.) {\bf 70} (1970) 10.
\bibitem{D}
P. Danielewicz, {Ann. Phys.} (N. Y.) {\bf 152} (1984) 239
\bibitem{Landsmann}N. P. Landsmann, Phys. Rev. Lett. {\bf 60} (1988)
1990; Ann. Phys. {\bf 186} (1988) 141; N. P. Landsmann and Ch.G. van Weert,
Phys. Rep. {\bf 145} (1987) 141.
\bibitem{DB}P. Danielewicz, G. Bertsch, Nucl. Phys. {\bf A533} (1991) 712.
\bibitem{BM}
W. Botermans and R. Malfliet, Phys. Rep. {\bf 198} (1990) 115.
\bibitem{HFN}M. Herrmann, B. L. Friman, W. N\"orenberg,
Nucl. Phys. {\bf A560} (1993) 411.
\bibitem{PH} P. A. Henning, Phys. Rep. {\bf C 253} (1995) 235;
Nucl. Phys. {\bf A 582} (1995) 633.
\bibitem{QH} E. Quack, P. A. Henning, GSI-95-29; Phys. Rev. Lett. in
print; GSI-95-42.
\bibitem{Weinhold} W. Weinhold, Diploma thesis, GSI 1995.
\bibitem{KV} J. Knoll and D. N. Voskresensky, Ann. Phys {\bf 249} (1996)
532; a condensed account of this work is published in
{Phys. Lett.}  {\bf B 351} (1995) 43.
\bibitem{Baym}
G. Baym, Phys. Rev. {\bf 127} (1962) 1391.
\bibitem{CGreiner}C. Greiner, K. Wagner, P.G. Reinhard,
  Phys.Rev. {\bf C49} (1994) 1693. 
\bibitem{LP} {E. M. Lifshitz and L. P. Pitaevskii}, ''Physical
Kinetics'' Nauka, 1979; Pergamon press, 1981.
\bibitem{LPM}
L. D. Landau and I. Pomeranchuk, Dokl. Akad. Nauk SSSR {\bf 92}
(1953) 553, 735; also in Collected Papers of Landau, ed. Ter Haar
(Gordon \& Breach, 1965) papers 75 - 77;
A. B. Migdal, Phys. Rev. {\bf 103}, (1956)1811;
Sov. Phys. JETP {\bf 5} (1957) 527. 
\bibitem{DGK} {J. Knoll and C. Guet}, {\it Nucl. Phys.} {\bf A494}
(1989) 334; {M. Durand and J. Knoll}, {\it Nucl. Phys.} {\bf
A496} (1989) 539;
{J. Knoll and R. Lenk}, {\it Nucl. Phys.} {\bf
A 561} (1993) 501.
\bibitem{Luttinger}
J. M. Luttinger and J. C. Ward, Phys. Rev. {\bf 118} (1960) 1417.
\bibitem{Cornwall} J.M. Cornwall, R. Jackiw and E. Tomboulis,
Phys. Rev. {\bf D 48} (1974) 2428.
\bibitem{IKV1}Yu. B. Ivanov, J. Knoll and D. N. Voskresenski,
  GSI-preprint-98-34, hep-ph/9807351.
\bibitem{KaBaym} G. Baym and L.P. Kadanoff, Phys. Rev. {\bf 124} (1961) 287.
\bibitem{KBBM}L.P. Kadanoff and G. Baym, ref. \cite{SKBK};
W. Botermans and R. Malfliet, Phys. Rep. {\bf 198} (1990) 115.
\bibitem{MLipS} R. Malfiet, Nucl. Phys. {\bf A 545} (1992) 2, and
Phys. Rev. {\bf B97} (1998) 11027;
 V. Spicka and P. Lipavsky, Phys. Rev. Lett.
{\bf 73} (1994) 3439;  Phys. Rev. {\bf B52} (1995) 14615.
\bibitem{IKV2}Yu. B. Ivanov, J. Knoll and D. N. Voskresenski, to be published.
\bibitem{BraatenPisarski}R. D. Pisarski, Nucl. Phys. A 525 (1991)
  175c; E. Braaten, Nucl. Phys. (Proc. Suppl.) {\bf B 23} (1991) 351.
\bibitem{Blaizot} J.P. Blaizot and E. Iancu, Nucl. Phys. {\bf B 390} (1993)
589, Phys. Rev. Lett. {\bf 70} (1993) 3376.
\bibitem{Jackiw93} R. Jackiw, V.P. Nair, Phys. Rev. {\bf D 48} (1993) 4991.
\end{thebibliography}
\end{document}